\documentclass[aps,prd,showkeys,showpacs,amssymb,cite,
amsfonts,epsf,preprintnumbers,nofootinbib,superscriptaddress]{revtex4}

\usepackage[dvips]{graphicx}
\usepackage{bm,latexsym,amsmath,amssymb,amsfonts,color}
\def\bi#1{\hbox{\boldmath{$#1$}}}

\begin{document}
\title{  Weak Lensing Effect on CMB in the Presence of a Dipole Anisotropy }

\author{Abhineet Agarwal} 
\email{abuwdu123@gmail.com}
\affiliation{Centre for Theoretical Physics,\\
 Jamia Millia Islamia, New Delhi-110025,India.}
   \author{Naveen K. Singh} 
\email{naveen.nkumars@gmail.com}
\affiliation{School of Physics and Astronomy,\\
 Sun Yat-Sen University Zhuhai,  2 Daxue Rd, Tangjia, Zhuhai, China.}
  
\author{Pankaj Jain} 
\email{pkjain@iitk.ac.in}
\affiliation{Department of Physics,\\
  Indian Institute of Technology, Kanpur 208016, India.}
 
\author{Prabhakar Tiwari} 
\email{tiwari.lko@gmail.com}
\affiliation{National Astronomical Observatories,\\
  CAS, Beijing 100012, China.}

\begin{abstract}
 We investigate weak lensing effect on cosmic microwave background (CMB) in the presence of dipole anisotropy. The approach of flat-sky approximation is considered. We determine the functions $\sigma_0^2$ and $\sigma_2^2$ that 
appear in expressions of the lensed CMB power spectrum in the presence of
 a dipole anisotropy. We determine the correction to B-mode power spectrum which is found to be appreciable at low multipoles ($l$).  However, the temperature and E-mode power spectrum are not altered significantly.
\\

\end{abstract}

\vspace{10mm}

\maketitle
 \section{Introduction} In unveiling the dynamics and contents of the universe, cosmic microwave background  radiation has been one of the essential tools. CMB constrains cosmological parameters, and it provides a path in developing modern cosmology. The precision level of
 the experiments for the CMB spectrum and hence the accuracy of prediction
 of a theoretical model have been increasing year by year. The CMB anisotropy 
 can be determined by the Physics of last scattering surface (LSS)
 and the medium effects due to the propagation of photons from LSS to
the observer. The primary contribution to the anisotropy is due to the strength of the metric perturbations at LSS. Some other effects such as ISW effect, reionization, quadrupole distributions give additional modifications. Lensing is an effect which is associated with the geodesics of photons from LSS to us. It can be considered as a secondary contribution in modifying CMB anisotropy, and it must be taken into account to reliably predict the CMB signal.
  The lensing gives a very small contribution on the reionization and the ISW signal as these are associated with the large scale. 
However, on degree-scale of primary acoustic peaks, weak lensing effect is expected at the order of arc minute 
on the CMB spectrum \cite{Lewis:2006fu}.  \\

In this paper, we are interested in the weak lensing effect on the CMB spectrum
arising due to a dipole distribution in the Newtonian potential. The unexpected observation of anisotropy at large scale leads us to study the implications of such model. In WMAP data, some of the anomalies were discovered \cite{Bennett:2012zja}. 
These anomalies include the alignment between low multipoles \cite{Tegmark:2003ve,Copi:2003kt, Land:2005ad}, large cold spot in the southern hemisphere \cite{Vielva:2003et, Mukherjee:2004in,Cruz:2004ce}, hemisphere power asymmetry \cite{Eriksen:2003db}. The Planck data has further confirmed the existence of such anomalies \cite{Ade:2013nlj}. According to Planck data, the dipole modulation of power lies in the range $l=2-600$ at $1.1-3.5 \sigma$. 
In addition, it confirmed octopole-quadrupole alignment at $98\%$ level and a power deficit below $l=40$. Other observables, such as, radio polarization from radio galaxies and optical polarization of quasars also indicate anisotropy at the large scale. Remarkably, 
orientations of dipole axis of radio polarization \cite{Jain:1998kf}, dipole, quadrupole and octopole axes of the  CMB \cite{Ralston:2003pf} and the direction of alignment of the pattern of  two-point correlation in the optical polarization \cite{Jain:2003sg} all point towards the Virgo direction. The model with dipole anisotropy we study in this paper 
is proposed by Gordon \cite{Gordon:2005ai}. It has also been studied in  Refs. \cite{Erickcek:2008sm,Liddle:2013czu,Ghosh:2013blq}. Furthermore, the CMB spectrum  have been investigated in Refs. \cite{Kothari:2015tqa,Ghosh:2015qta, Ghosh:2018apx} in this model. In this paper,  we study weak lensing effect on the 
CMB spectrum with the flat-sky approximation \cite{Zaldarriaga:1998ar}.  In Sec. (\ref{Sec2}), we generalize the two-point correction function due to the weak lensing with the dipole anisotropy. 
In this section, we also calculate the
expectation value of a term which arises due to the isotropic part of lensing potential.
In Sec. (\ref{Sec3}), we estimate the modification to this term in  presence of the dipole anisotropy.  In Sec. (\ref{Sec4}), we obtain expressions for CMB power spectrum. In Sec. (\ref{Sec5}), we conclude.

 \section{Gravitational Lensing Effect on Two-Point Correlation Functions in the Presence of Dipole Anisotropy}\label{Sec2}
The observed CMB spectrum is different from that produced at LSS due to
the lensing effect.
These are related as \cite{Zaldarriaga:1998ar} 
  \begin{eqnarray}
   X(\bi \theta) &=& \tilde{X}(\bi \theta + \delta \bi \theta) \nonumber \\
             &=&  \frac{1}{(2 \pi)^2} \int d^2 {\bi l} e^{i {\bi l}.(\bi \theta+\delta \bi \theta)} \tilde{X}({\bi l}) ,   
  \end{eqnarray}
where, $X( \bi \theta)$ is  the observed physical quantity such as the temperature $T( \bi \theta)$, the stokes parameters $Q( \bi \theta)$ and
$U(\bi \theta)$ etc.,
whereas $\tilde{X}(\bi \theta + \delta \bi \theta)$ is the corresponding
physical quantity produced at LSS. Writing these variables in the explicit form, 
we have,
\begin{eqnarray}
T({\bi \theta})&=&\tilde T( \bi \theta+\delta \bi \theta) \nonumber \\
	 &=&\frac{1}{(2\pi)^{2}}\int d^2 {\bi l}\ 
e^{i{\bi l}\cdot ({\bi \theta}+ \delta {\bi \theta})}\tilde T({\bi l}), \label{temp1}\\
Q({ \bi \theta})&=&\tilde Q(\bi \theta+\delta \bi \theta) \nonumber \\
	 &=&\frac{1}{(2\pi)^{2}}\int d^2 {\bi l}\ 
e^{i{\bi l}\cdot ({\bi \theta}+ \delta {\bi \theta})}\tilde Q({\bi l}), \nonumber \\
 &=&\frac{1}{(2\pi)^{2}}\int d^2 {\bi l}\ 
e^{i{l}\cdot ({\bi \theta}+ \delta {\bi \theta})}\Big[\tilde E({\bi l}) \cos(2 \phi_{\bi l}) - \tilde B({\bi l})\sin(2 \phi_{\bi l}) \Big] \label{Q1},\\
U({ \bi \theta})&=&\tilde U(\bi \theta+\delta \bi \theta) \nonumber \\
	 &=&\frac{1}{(2\pi)^{2}}\int d^2 {\bi l}\ 
e^{i{\bi l}\cdot ({\bi \theta}+ \delta {\bi \theta})}\tilde U({\bi l}) \nonumber \\
 &=&\frac{1}{(2\pi)^{2}}\int d^2 {\bi l}\ 
e^{i{\bi l}\cdot ({\bi \theta}+ \delta {\bi \theta})}\Big[\tilde E({\bi l}) \sin(2 \phi_{\bi l}) + \tilde B({\bi l})\cos(2 \phi_{\bi l}) \Big] , \label{U1}
\end{eqnarray}
where, the standard definitions of $Q({\bi l})$ and  $U({\bi l})$ are used. Using above Eqs. (\ref{temp1}), (\ref{Q1}) and (\ref{U1}), we can compute 
correlation function of such quantities at two points $\theta_1$ and $\theta_2$. In general, these two points can be anywhere 
on the two dimensional plane in small scale limit formalism. However for simplicity, we rotate our frame such that these two points
fall on $x-$axis. We can also shift $\theta_1$ at $\theta_1=0$ or $\theta_2$ at $\theta_2=0$ keeping other point at the separation  of $\theta$.
Now, the correlation function corresponding to the temperature is given by,
\begin{eqnarray}
 C_{T}(\theta) &=& <T({\bi \theta_1}) T({\bi \theta_2)}> \nonumber \\
              &=& <\tilde T({\bi \theta_1}+ \delta {\bi \theta_1}) \tilde T({\bi \theta_2} + \delta {\bi \theta_2})> \nonumber \\
              &=& \int \frac{d^2 {\bi l}}{(2 \pi)^2} e^{i {l} \theta \cos(\phi_{l})} < e^{i {\bi l} \cdot (\delta {\bi \theta_1} - \delta {\bi \theta_2}) }> C_{\tilde T l},\label{corrTemp}
\end{eqnarray}
where, we have used
\begin{eqnarray}
 <\tilde Y({\bi l})\tilde Y({\bi l}')> = (2 \pi)^2 \tilde C_{Y {\bi l}} \delta^2 ({\bi l}-{\bi l}').
\end{eqnarray}
The vector quantity $\bi l$ which is conjugate to $ \bi \theta$ makes the angle $\phi_{\bi l}$ with the $x-$axis. In the similar way, we obtain the following 
relation for $C_{Q}(\theta)$ and $C_{U}(\theta)$,
\begin{eqnarray}
C_Q({\theta})&=&\int {d^2{\bi l}\over (2\pi)^2}\  
e^{il\theta \cos \phi_{l}}
\langle e^{i{\bi l}\cdot (\delta{\bi \theta_1} -\delta{\bi
\theta}_{2})}\ \rangle\ \ [C_{\tilde{E} l} \cos^2(2\phi_{\bi l})
+C_{\tilde{B} l} \sin^2(2\phi_{\bi l})]\nonumber \\
C_U({\theta})&=&\int {d^2{\bi l}\over (2\pi)^2}\  
e^{il\theta \cos \phi_{l}}
\langle e^{i{\bi l}\cdot (\delta{\bi \theta_1} -\delta{\bi
\theta}_{2})}\ \rangle\ \ [C_{\tilde{ E}l} \sin^2(2\phi_{\bi l})
+C_{\tilde{ B}l} \cos^2(2\phi_{\bi l})].
\label{corrlens}
\end{eqnarray}
The quantity $\langle e^{i{\bi l}\cdot (\delta{\bi \theta_1} -\delta{\bi
\theta}_{2})}\ \rangle $ is important since the weak lensing effect with the dipole anisotropy is involved here. We compute this quantity following Ref. \cite{Zaldarriaga:1998ar} by adding the dipole matter distribution.  Let us define the deflection angles as $\alpha=\delta \theta_1$ and $\alpha'=\delta \theta_2$. The deflection angle is given by,

\begin{eqnarray}
 \alpha = \nabla_{\hat{n}} \psi(\hat{n}),
\end{eqnarray}
where, $\nabla_{\hat{n}}$ is angular derivative, $\hat{n}$ is direction of observation and $\psi(\hat{n})$ is lensing potential and it is connected to gravitational potential ${\bi \psi}(\chi \hat{n}, \eta_0-\chi)$ via,

\begin{eqnarray}
 \psi(\hat{\eta}) \equiv -2 \int^{\chi^*}_{0} d \chi \frac{f_K(\chi^*-\chi)}{f_K(\chi^*)f_K(\chi)} {\bi \psi}(\chi \hat{n}, \eta_0-\chi),
\end{eqnarray}
where, $\chi^*$ is conformal distance of LSS from us. In our case, the gravitational potential has two parts, the homogeneous part ${\bi \psi}^h $and
part ${\bi \psi}^d$ corresponding to dipole distribution: 
\begin{eqnarray}
 {\bi \psi} = {\bi \psi}^h + {\bi \psi}^d.
\end{eqnarray}
Expanding ${\bi \psi}^h$ by Fourier expansion, we have 
\begin{eqnarray}
  {\bi \psi}(\theta) = \frac{1}{(2 \pi)^2}\int d^2 {\bi l} e^ {i {\bi l} \cdot \bi \theta} {\bi \psi}^h_{\bi l} +  {\bi \psi}^d .
\end{eqnarray}
Now we calculate the correlation tensor,
\begin{eqnarray}
 \langle \alpha_i \alpha'_j \rangle = \langle \nabla_{i} \psi  \nabla_{j} \psi'\rangle. \label{eqn_cor1}
\end{eqnarray}
Using above Eq. (\ref{eqn_cor1}), we obtain
\begin{eqnarray}
 \langle \alpha_i \alpha'_j \rangle &=&  \langle \nabla_{i} { \psi}  \nabla_{j} { \psi}'\rangle, \nonumber \\
 &=& \langle \nabla_{i} { \psi^h}  \nabla_{j} { \psi'{^h}}\rangle +
 \langle \nabla_{i} { \psi^d}  \nabla_{j} { \psi'^{d } }\rangle, \nonumber \\
 &=&  \int \frac{d^2 {\bi l}}{(2 \pi)^2}  {\bi l}_i {\bi l}_j e^{i {\bi l} \cdot {\bi \theta}} C_{ l}^{  \psi  \psi'}
 + \langle \nabla_{i} { \psi^d}  \nabla_{j} { \psi'^{d } }\rangle
  , \label{eqn_cor2}
\end{eqnarray}
where, $\langle { \psi}^h_{\bi l} { \psi}^{h *}_{\bi l'}\rangle= (2 \pi)^2 \delta^2({\bi l}-{\bi l'}) C_{ l}^{  \psi  \psi'} $
is used. In Eq. (\ref{eqn_cor2}), we denote the first term which is the homogeneous part by $\langle \alpha^h_i  \alpha'^h_j \rangle$. The correlation function
$\langle \alpha^h_i  \alpha'^h_j  \rangle$ should be proportional to $\delta_{ij}$ and the trace-free tensor given by 
$\hat{r}_{<i}\hat{r}_{j>}= \hat{r}_{i}\hat{r}_{j}- \frac{1}{2}\delta_{ij}$. Using this fact, the first homogeneous part simplifies as
\begin{eqnarray}
 \int \frac{d^2 {\bi l}}{(2 \pi)^2}  {\bi l}_i {\bi l}_j e^{i {\bi l} \cdot {\bi \theta}} C_{ l}^{  \psi  \psi'} = 
 \frac{1}{2} C_{g l } \delta_{ij}- C_{ g l, 2} \hat{r}_{<i}\hat{r}_{j>},
\end{eqnarray}
where,
\begin{eqnarray}
 C_{g l }(\theta) = \int {\frac{d l}{2 \pi} l^3 }  C_{ l}^{  \psi  \psi'} J_{0}( l \theta), \\
  C_{g l,2 }(\theta) = \int {\frac{d l}{2 \pi} l^3 }  C_{ l}^{  \psi  \psi'} J_{2}( l \theta) .
\end{eqnarray}

Now we come to the term $\langle e^{i {\bi l}\cdot (\alpha -\alpha')} \rangle $ in which we are interested in order to understand the lensing effects
including dipole matter distribution. It can be shown that \cite{Lewis:2006fu}
\begin{eqnarray}
 \langle e^{i {\bi l}\cdot (\alpha -\alpha')} \rangle = e^{-\frac{1}{2}\langle[l \cdot (\alpha -\alpha')]^2\rangle}.
\end{eqnarray}
Expanding the term $\langle[l \cdot (\alpha -\alpha')]^2\rangle$, we have
\begin{eqnarray}
 \langle[l \cdot (\alpha -\alpha')]^2\rangle &=& l^i l^j \langle (\alpha -\alpha')_i (\alpha -\alpha')_j\rangle \nonumber \\
 &=&l^i l^j \Big[ \langle \alpha_i \alpha_j \rangle + \langle \alpha_i' \alpha_j' \rangle - 2 \langle \alpha_i \alpha_j' \rangle \Big] \nonumber \\
 &=& l^i l^j \Big[ \langle \alpha_i \alpha_j \rangle + \langle \alpha_i' \alpha_j' \rangle - 2 \langle \alpha_i \alpha_j' \rangle \Big]^h + l^i l^j\Big[ \langle \alpha_i \alpha_j \rangle + \langle \alpha_i' \alpha_j' \rangle - 2 \langle \alpha_i \alpha_j' \rangle \Big]^d.
\end{eqnarray}

We obtain the terms due to homogeneous part as well as dipole part, since $\langle \alpha_i \alpha_j \rangle$, 
$\langle \alpha_i' \alpha_j' \rangle$ and $\langle \alpha_i \alpha_j' \rangle$ all have two parts as in Eq. (\ref{eqn_cor2}). Homogeneous part
becomes 
\begin{eqnarray}
\langle[l \cdot (\alpha -\alpha')]^2\rangle^h &=&  l^2  \Big[ C_{g l}(0) - C_{g l }(\theta)  +  \cos(2 \phi_l)  C_{g l,2 }(\theta)\Big], \nonumber \\
&=& l^2 \Big[ \sigma_0^2(\theta) + \sigma_2^2(\theta) \cos(2 \phi_l)\Big],
\end{eqnarray}
where, we define $\sigma_0^2(\theta) = C_{g l}(0) - C_{g l }(\theta)  $ and $\sigma_2^2(\theta) = C_{gl,2}$. These notations, $\sigma_0^2(\theta)$ and $\sigma_2^2(\theta)$,  have been used in the literature \cite{Zaldarriaga:1998ar}.
\section{Dipole Anisotropy Correction}\label{Sec3}
In this section, we estimate first the correction to $\langle[l \cdot (\alpha -\alpha')]^2\rangle$ due to the dipole distribution.
In Eq. (\ref{eqn_cor2}), the additional second term is due to the dipole. To calculate it, we expand $\psi^d$ in $3$-dimensional Fourier space as follows \cite{Yu:2009bz},
\begin{eqnarray}
 \psi^d(\eta^*,\hat{n}) = -2 \int^{\chi^*}_{0} d \chi_1 \frac{f_K(\chi^* -\chi_1)}{f_K(\chi^*)f_K(\chi_1)} \int \frac{d^3k}{(2 \pi)^3} {\bi \psi}^d({\bi k})e^{i k.(\eta_0 -\eta_1)\hat{n}} T_{\psi}(\eta_1,k). \label{eqn_psidk}
\end{eqnarray}
The dipole distribution is defined by the coefficient ${\bi \psi}^d_{\bi k}$ which is given by \cite{Ghosh:2013blq},
\begin{eqnarray}
 {\bi \psi}^d_{\bi k} = \frac{\beta}{2 i} (2 \pi)^3\Big[\delta^3({\bi k} - \kappa_s \hat{x})- \delta^3({\bi k} + \kappa_s \hat{x})\Big],
\end{eqnarray}
where, $\beta$ and $\kappa_s$ are two parameters which define the dipole. We consider the alignment of dipole in $x-$direction and we keep $ \psi^d$ in generalized form of 3D. 
Now, we redefine the momentum $k$ in terms of  $l$, by $\vec{l}= \chi^* \vec{k}$, $\int d^3 k {\bi\psi}^d(k) \rightarrow \int d^3l {\bi \psi}^d(l)$ and $\frac{\eta_0-\eta}{\chi^*} = \frac{\chi}{\chi^*} =\omega$. In this definition, 
\begin{eqnarray}
 \frac{{\bi \psi}^d(k)}{\chi^{*3}} &=& \frac{\beta}{2 i} \frac{(2 \pi)^3}{\chi^{*3}}\Big[\delta^3({\bi k} - \kappa_s \hat{x})- \delta^3({\bi k} + \kappa_s \hat{x})\Big] \nonumber \\
 &=& \frac{\beta}{2 i} (2 \pi)^3 \Big[\delta^3( \chi^* {\bi k} - \chi^*\kappa_s \hat{x})- \delta^3(\chi^*{\bi k} + \chi^*\kappa_s \hat{x})\Big],\nonumber \\
 &=&\frac{\beta}{2 i} (2 \pi)^3\Big[\delta^3({\bi l} - \kappa \hat{x})- \delta^3({\bi l} + \kappa \hat{x})\Big] = \psi^d(l),
\end{eqnarray}
where, $\kappa= \chi* \kappa_s$. Under such definition, Eq. (\ref{eqn_psidk}) can be written as,
\begin{eqnarray}
 \psi^d(\eta^*,\hat{n}) = -2 \int^{\chi*}_{0} d \chi_1 \frac{f_K(\chi^* -\chi_1)}{f_K(\chi*)f_K(\chi_1)} \int \frac{d^3l}{(2 \pi)^3} \psi^d(l)e^{i {\bi l}.\hat{n} \omega}  T_{\psi}(\eta_1,l).
\end{eqnarray}
We assume  that the transfer function is nearly unity for the large scale and  set it as unity in the further calculation. We can now simplify the term  $ \langle \nabla_{i} { \psi^d}  \nabla_{j} { \psi'^{d } }\rangle$ as follows,
\begin{eqnarray}
 <\nabla_i \psi^d(\eta^*, \hat{n_1}) \nabla_j \psi^{*d}(\eta^*, \hat{n_2})>
 = 4 \int^{\chi^*}_{0}\int^{\chi^*}_0 d\chi_1 d \chi_2 \frac{(\chi^* -\chi_1)}{(\chi^* \chi_1)}\frac{(\chi^* -\chi_2)}{(\chi^* \chi_2)} \times \nonumber \\  \int \int d^3 l d^3 l' \omega_1 \omega_2 l_i l_j' \left(\frac{\beta^2}{4}\right)\Big[\delta^3(\vec{l}-\kappa \hat{x}) \delta^3(\vec{l}-\kappa \hat{x})-\delta^3(\vec{l}-\kappa \hat{x})\delta^3(\vec{l}+\kappa \hat{x}) \nonumber \\ -\delta^3(\vec{l}+\kappa \hat{x})\delta^3(\vec{l}-\kappa \hat{x})+\delta^3(\vec{l}+\kappa \hat{x})\delta^3(\vec{l}+\kappa \hat{x})\Big]\times \nonumber \\
 e^{i \hat{l}. \hat{n_1} \omega_1}e^{-i \hat{l'}. \hat{n_2} \omega_2}.
\end{eqnarray}
Here, $\eta$ is the conformal time at the point of observation.  $\nabla_i$ has two components corresponding to the directions $x$ and $y$ and in the flat sky approximation equivalent to $\frac{\partial}{\ \partial {n_x}}$ and  $\frac{\partial}{\ \partial {n_y}}$ respectively. Here $n_x$ and $n_y$ are  x and y components of $\hat{n}$, the 
unit vector in the direction of observation from us. The two points of observation could be anywhere on the sphere, however, we consider these points on the great circle parallel to $x$-axis.   The unit direction $\hat{n}$, which is
given by
\begin{eqnarray}
 \hat{n} = \sin{\theta}\cos{\phi}\hat{x} + \sin{\theta}\sin{\phi}\hat{y}+ \cos{\theta}\hat{z} \label{eqn_rad}
\end{eqnarray}
becomes $\sin{\theta}\hat{x} + \cos{\theta}\hat{z}$ ($\phi =0$) in this case. The momentum integrals contains four terms, we calculate these one by one. The first term is given by,
\begin{eqnarray}
 \frac{\beta^2}{4} \int \int d^3 l d^3 l' \omega_1 \omega_2 l_i l_j' \Big[\delta^3(\vec{l}-\kappa \hat{x}) \delta^3(\vec{l'}-\kappa \hat{x})\Big] e^{i \hat{l}. \hat{n_1} \omega_1}e^{-i \hat{l'}. \hat{n_2} \omega_2} \nonumber\\
 = \frac{\beta^2}{4} \omega_1 \omega_2 \kappa^2 \delta_i^x \delta_j^x e^{i \kappa(\omega_1 \sin \theta_1  - \omega_2 \sin \theta_2 ) },
\end{eqnarray}
similarly second, third and last terms can be written as, 
\begin{eqnarray}
 -\frac{\beta^2}{4} \int \int d^3 l d^3 l' \omega_1 \omega_2 l_i l_j' \Big[\delta^3(\vec{l}-\kappa \hat{x}) \delta^3(\vec{l'}+\kappa \hat{x})\Big] e^{i \hat{l}. \hat{n_1} \omega_1}e^{-i \hat{l'}. \hat{n_2} \omega_2} \nonumber\\
 = \frac{\beta^2}{4} \omega_1 \omega_2 \kappa^2 \delta_i^x \delta_j^x e^{i \kappa( \omega_1 \sin \theta_1  +  \omega_2 \sin\theta_2) },
\end{eqnarray}

\begin{eqnarray}
 -\frac{\beta^2}{4} \int \int d^3 l d^3 l' \omega_1 \omega_2 l_i l_j' \Big[\delta^3(\vec{l}+\kappa \hat{x}) \delta^3(\vec{l'}-\kappa \hat{x})\Big] e^{i \hat{l}. \hat{n_1} \omega_1}e^{-i \hat{l'}. \hat{n_2} \omega_2} \nonumber\\
 = \frac{\beta^2}{4} \omega_1 \omega_2 \kappa^2 \delta_i^x \delta_j^x e^{-i \kappa(\omega_1 \sin \theta_1  + \omega_2 \sin \theta_2 ) },
\end{eqnarray}
and 
\begin{eqnarray}
 \frac{\beta^2}{4} \int \int d^3 l d^3 l' \omega_1 \omega_2 l_i l_j' \Big[\delta^3(\vec{l}+\kappa \hat{x}) \delta^3(\vec{l'}+\kappa \hat{x})\Big] e^{i \hat{l}. \hat{n_1} \omega_1}e^{-i \hat{l'}. \hat{n_2} \omega_2} \nonumber\\
 = \frac{\beta^2}{4} \omega_1 \omega_2 \kappa^2 \delta_i^x \delta_j^x e^{-i \kappa(\omega_1 \sin \theta_1  - \omega_2 \sin\theta_2 ) },
\end{eqnarray}
respectively. Summing all the terms, we obtain the total contribution, which is given by,
\begin{eqnarray}
  \beta^2 \omega_1 \omega_2 \kappa^2 \delta^x_i \delta^x_j \cos{\left(\kappa \omega_1  \sin\theta_1 \right)}\cos{\left(\kappa  \omega_2  \sin\theta_2 \right)}
\end{eqnarray}
As already mentioned, we may shift one point to the pole. Here we shift the one point to the pole by making $\theta_2 \rightarrow 0$. We can also now redefine $\theta_1$ as only  $\theta$, therefore, above term can be written as,
\begin{eqnarray}
 \beta^2 \omega_1 \omega_2 \kappa^2 \delta^x_i \delta^x_j \cos{\left(\kappa \omega_1 \sin\theta \right)}.
 \end{eqnarray}
Now we integrate this term with respect to conformal time by recalling  $\omega= \frac{\chi}{\chi^*}$, and we obtain,

\begin{eqnarray}
 <\nabla_i \psi^d(\eta^*, \hat{n_1}) \nabla_j \psi^{*d}(\eta^*, \hat{n_2})>
 = 2 \beta^2 \delta_i^x \delta_j^x  \Big[\frac{1- \cos{[\kappa \sin\theta]}}{\sin^2\theta}\Big].
\end{eqnarray}

In the notation of $\alpha_i$, 

\begin{eqnarray}
 <\alpha_i^d \alpha_j'^d> = 2 \beta^2 \delta_i^x \delta_j^x  \Big[\frac{1- \cos{[\kappa  \sin \theta]}}{\sin^2 \theta}\Big],
\end{eqnarray}
and so,
\begin{eqnarray}
 <\alpha_i^d \alpha_j^d> =  <\alpha_i'^d \alpha_j'^d> = \lim_{\theta \to 0} <\alpha_i^d \alpha_j'^d> =  \beta^2 \kappa^2 \delta_i^x \delta_j^x  .
\end{eqnarray}
 Our motivation is to calculate $<|\vec{l}. (\vec{\alpha_i}-\vec{\alpha_i}')|>^{d}$. Keeping all quantities in hand, we can simplify this quantity as, 
\begin{eqnarray}
 <|\vec{l}. (\vec{\alpha}-\vec{\alpha}')|^2>^{d} = l^i l^j \Big[<\alpha_i^d \alpha_j^d> + <\alpha_i'^d \alpha_j'^d> - 2<\alpha_i^d \alpha_j'^d>\Big] \nonumber \\
 = \beta^2 l^2 \left(1 +  \cos{2 \phi_l}\right) \Big[\kappa^2 - 2 \frac{(1 - \cos{[\kappa  \sin \theta]})}{\sin^2 \theta }\Big]. \label{ave_term}
\end{eqnarray}
\section{CMB Power spectrum}\label{Sec4}
We observe that the dipole anisotropy modifies the quantities $C_{gl}(0) -C_{gl}(\theta)$ ($\equiv \sigma_0^2$) and $C_{gl,2}$ ($ \equiv \sigma_2^2$) with equal weight. Therefore,$\langle e^{i {\bi l}\cdot (\alpha -\alpha')} \rangle $ turns out to be,
\begin{eqnarray}
 \langle e^{i {\bi l}\cdot (\alpha -\alpha')} \rangle = \exp\{-\frac{l^2}{2} \big[\sigma_{0T}^2 + \cos(2 \phi_l) \sigma_{2T}^2\big]\}, \label{phaseterm}
\end{eqnarray}
where, we define new quantities $\sigma_{0T}^2$ and $\sigma_{2T}^2$ as,
\begin{eqnarray}
 \sigma_{0T}^2 &=&  C_{gl}(0) -C_{gl}(\theta)  + \beta^2 \Big[\kappa^2 - 2 \frac{(1 - \cos{[\kappa  \sin \theta]})}{\sin^2 \theta }\Big], \nonumber \\
 &=& \sigma_0^2 + \sigma_d^2,
 \label{eqn1}
\end{eqnarray}
and,
\begin{eqnarray}
 \sigma_{2T}^2 &=&  C_{gl,2}(\theta)  + \beta^2 \Big[\kappa^2 - 2 \frac{(1 - \cos{[\kappa  \sin \theta]})}{\sin^2 \theta }\Big]\nonumber \\
 &=& \sigma_2^2 + \sigma_d^2,
 \label{eqn2}
 \end{eqnarray}
 where, $\sigma_d^2$ is given by
 \begin{eqnarray}
  \sigma_d^2 = \beta^2 \Big[\kappa^2 - 2 \frac{(1 - \cos{[\kappa  \sin \theta]})}{\sin^2 \theta }\Big].
  \end{eqnarray}
In Eq. (\ref{phaseterm}), we note that the term is similar as we get in the standard lensing. The only difference here is that all the effects of anisotropy are
absorbed in $\sigma_{0T}^2$ and $\sigma_{2T}^2$. This facilitates us in using further construction of standard weak lensing. Therefore, from Eqs. (\ref{corrTemp}) and (\ref{corrlens}), we obtain,
\begin{eqnarray}
C_T(\theta)&=&\int \frac{ l dl} { 2\pi}
\ C_{\tilde Tl}\ \Big[J_0(l\theta)[1-{l^2 \over 2}\sigma_{0T}^2(\theta) ]
+{l^2 \over 2}\sigma_{2T}^2(\theta)
J_2(l\theta)\Big] , \nonumber \\ 
C_{Q}(\theta)+C_{U}(\theta)&=&\int {l dl\over 2\pi}
\ (C_{\tilde El}+C_{\tilde Bl})\ 
\Big[ J_0(l\theta)[1-{l^2 \over 2}\sigma_{0T}^2(\theta) ]
+{l^2 \over 2} \sigma_{2T}^2(\theta)
J_2(l\theta)\Big] ,  \nonumber \\ 
C_{Q}(\theta)-C_{U}(\theta)&=&\int {l dl\over 2\pi}
\ (C_{\tilde El}-C_{\tilde Bl})\ 
\Big[ J_4(l\theta)[1-{l^2 \over 2}\sigma_{0T}^2(\theta) ]
+{l^2 \over 4} \sigma_{2T}^2(\theta)
[J_2(l\theta)+J_6(l\theta)]\Big].
\label{corrlens1}
\end{eqnarray}
We can now estimate the power spectrum in Fourier space from widely used definitions,
\begin{eqnarray}
C_{Tl}&=&2\pi \int_0^\pi \theta d\theta \ C_T(\theta) \ J_0(l\theta), \nonumber \\
C_{El}&=&2\pi \int_0^\pi \theta d\theta \
\Big[ [C_Q(\theta)+C_U(\theta)]\ J_0(l\theta) 
+ [C_Q(\theta)-C_U(\theta)]\ J_4(l\theta) \Big], \nonumber \\
C_{Bl}&=&2\pi \int_0^\pi \theta d\theta \
\Big[ [C_Q(\theta)+C_U(\theta)]\ J_0(l\theta) 
- [C_Q(\theta)-C_U(\theta)]\ J_4(l\theta) \Big].
\label{pow_spect}
\end{eqnarray}
From Eq. (\ref{corrlens1}) and (\ref{pow_spect}), we can the write expressions for
all power spectrum which reveal how modified power spectrum are different from the primordial
ones. These are as follows,
\begin{eqnarray}
C_{Tl}&=&C_{\tilde Tl}+{\cal W}_{1 l}^{l^\prime}\ C_{\tilde Tl^{\prime}},
\nonumber \\
C_{El}&=&C_{\tilde El}+{1 \over 2}[{\cal W}_{1 l}^{l^\prime}+{\cal W}_{2
l}^{l^\prime}]\  C_{\tilde El^{\prime}} + {1 \over 2}[{\cal W}_{1
l}^{l^\prime}-{\cal W}_{2 l}^{l^\prime}]\  C_{\tilde Bl^{\prime}} ,
\nonumber \\
C_{Bl}&=&C_{\tilde Bl}+{1 \over 2}[{\cal W}_{1 l}^{l^\prime}-{\cal W}_{2
l}^{l^\prime}]\  C_{\tilde El^{\prime}} + {1 \over 2}[{\cal W}_{1
l}^{l^\prime}+{\cal W}_{2 l}^{l^\prime}]\  C_{\tilde Bl^{\prime}} ,
\label{pow_spect1}
\end{eqnarray}
where we sum implicitly over all $l^{\prime}$ and  ${\cal W}^{l'}_{1 l}$ and ${\cal W}^{l'}_{2 l}$ are defined as,
\begin{eqnarray}
{\cal W}_{1 l}^{l^\prime}&=&{{l^\prime}^3 \over 2}\int_0^{\pi} \theta d\theta
\ J_0(l\theta)\
\Big[\sigma_{2T}^2(\theta)J_2(l^{\prime}\theta)
-\sigma_{0T}^2(\theta)J_0(l^{\prime}\theta)\Big],
\nonumber \\
{\cal W}_{2 l}^{l^\prime}&=&{{l^\prime}^3 \over 2}\int_0^{\pi} \theta d\theta
\ J_4(l\theta)\
\Big[{1 \over 2}\sigma_{2T}^2(\theta)
[J_2(l^{\prime}\theta)+J_6(l^{\prime}\theta)]
-\sigma_{0T}^2(\theta)J_4(l^{\prime}\theta)\Big].
\end{eqnarray}

\begin{figure}[h]
\begin{tabular}{cc}
\includegraphics[width=0.45\linewidth]{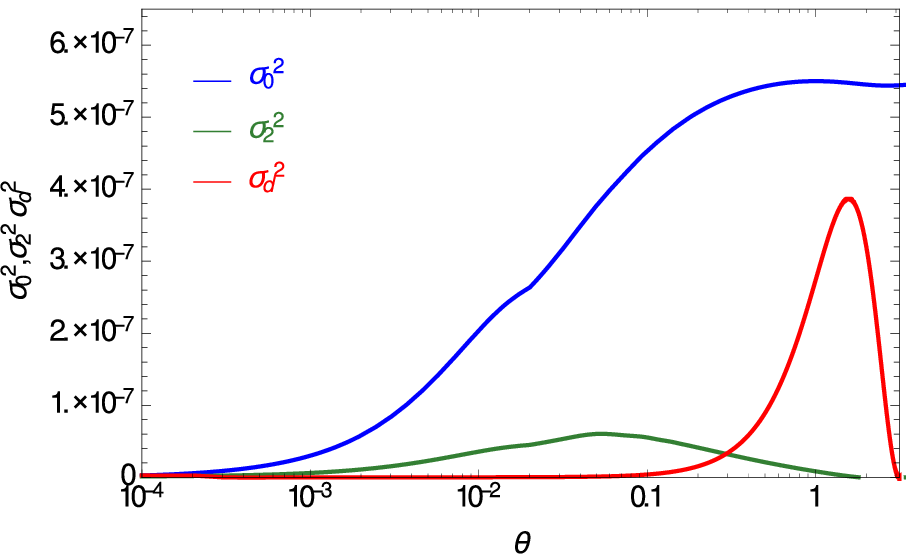}
  \includegraphics[width=0.44\linewidth]{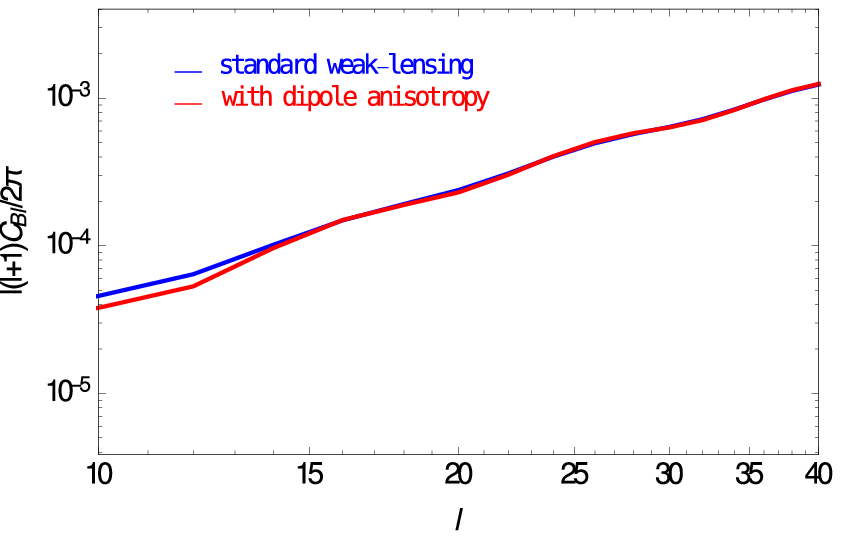}
\end{tabular}
\caption{ The left panel shows plot for $\sigma_0^2$, $ \sigma_2^2$ and $\sigma_d^2$.  On the right panel, blue curve is B-mode power spectrum for the standard weak lensing and red curve is for the weak lensing with the dipole distribution. }
\label{fig1}
\end{figure}

\begin{figure}[h]
\begin{tabular}{cc}
\includegraphics[width=.45\linewidth]{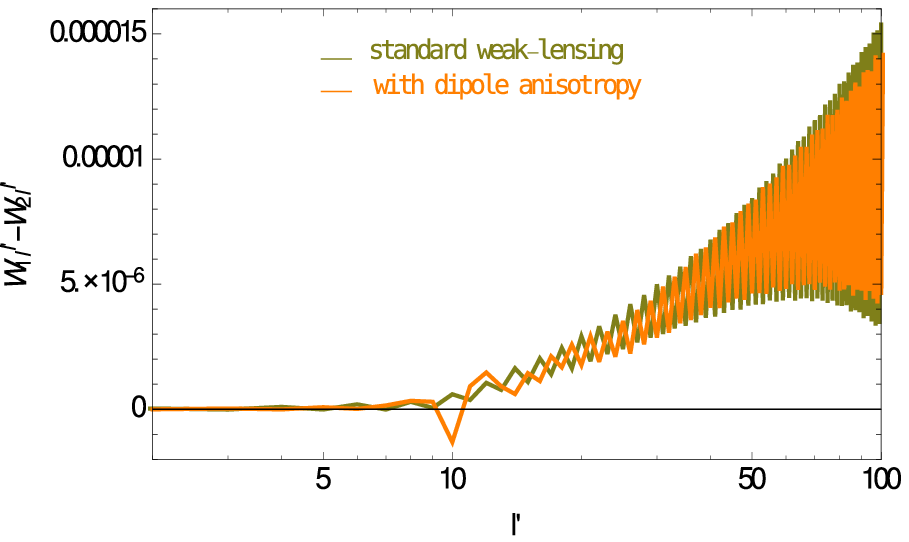}
   \includegraphics[width=0.45\linewidth]{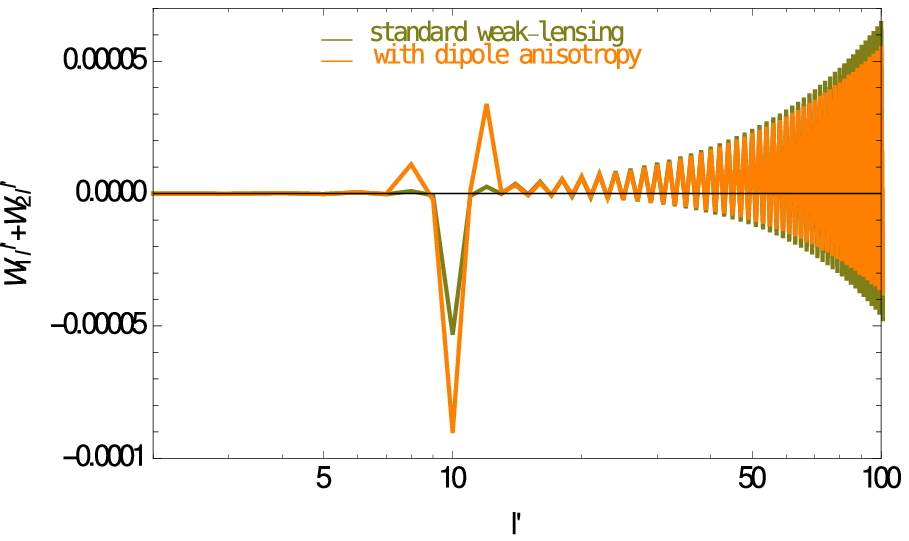}
\end{tabular}
\caption{ $W^{l'}_{1l}-W^{l'}_{2l}$ and $W^{l'}_{1l}+W^{l'}_{2l}$ are plotted with respect to $l'$ for $l=10$ in the left and right panel respectively. }
\label{fig2}
\end{figure}
\begin{table}[h!]
\caption{ fractional change in B-mode power spectrum from $l=20$ to $100$}
\label{tabfrac1}
\begin{center}
  \begin{tabular}{ | p{4cm} | p{2cm} | c | c | c | c | }
    \hline
    ``l'' & 20 & 40 & 60& 80 &100 \\ \hline
    \hspace{0.2cm}$l(l+1)C_{Bl}/2 \pi  $ (standard lensing) & $0.000238658$ & $0.00123632$ &  $ 0.002661348$& $0.004883204$ & $0.00772255$  \\ \hline
     $l(l+1) C_{Bl}/2 \pi  $ (lensing with dipole anisotropy) & $0.000230123$ & $0.0012477 $& $0.002653185$& $0.004874728$ & $0.007715724$   \\
    \hline
    fractional change in  $l(l+1) C_{Bl}/2 \pi  $ & 0.035762 &0.0092 & 0.003067 & 0.001736 & 0.000884  \\
    \hline
  \end{tabular}
\end{center}
\end{table}
\begin{figure}[h]
\begin{tabular}{c}
\includegraphics[width=.45\linewidth]{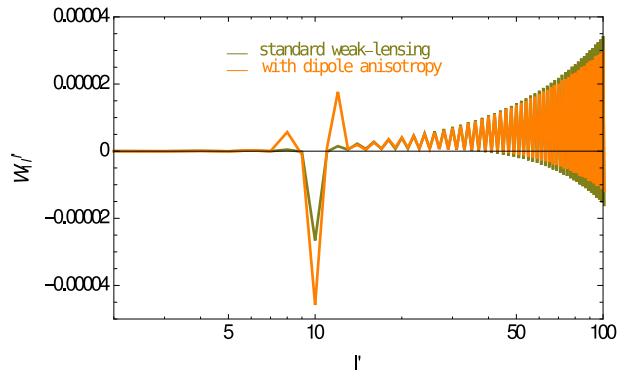}
\end{tabular}
\caption{ The plot of $W^{l'}_{1l}$  with respect to $l'$ for $l=10$.  }
\label{fig3}
\end{figure}
We note that in Eq. (\ref{pow_spect1}), the modified power spectrums are determined by functions $W_{11}^{l'}$, $W_{21}^{l'}$ and $W_{31}^{l'}$ those depend on $\sigma_{0T}^2$ and $\sigma_{2T}^2$. Since theoretically, the primordial 
$C_{\tilde{B} l}$ has nearly zero contribution, the correction to $C_{\tilde{E} l}$ is proportional to $W_{11}^{l'}+ W_{11}^{l'}$ and the only contribution to $C_{B l}$ comes from the correction which is proportional to $W_{11}^{l'}- W_{11}^{l'}$. 
In Fig. (\ref{fig1}), on the left panel, we plot $\sigma_0^2$, $\sigma_2^2$ which are due to standard weak lensing. The function $\sigma_d^2$ due to dipole distribution is also plotted. 
We considered $\beta=1$ and $\kappa = 0.046$ satisfying the constraint $|\kappa_s^3 \beta| \leq 1.26 \times 10^{-5} H_0^3$ or $|\kappa^3 \beta| \leq   10^{-4}$\cite{Ghosh:2013blq}, where, $H_0$ is Hubble constant. This provides with an estimate of maximum possible change that can
arise due to the dipole. We observe a reasonable value of $\sigma_d^2$ comparable to $\sigma_0^2$. The contribution of $\sigma_d^2$ appears in the B-mode spectrum. Dipole distribution marginally decreases its value due to the presence of function  $W_{11}^{l'}- W_{11}^{l'}$. 
In the right panel of Fig. (\ref{fig1}), we plot the $B$-mode power spectrum
including the dipole contribution. The standard weak lensing result is 
shown for comparison.
In plotting the lensed $C_{Bl}$, we use unlensed $\tilde C_{Bl}=0$ and $\tilde C_{El}$ from CAMB package. We restrict this curve for $l>10$ for which the 
 Flat sky approximation is expected to be reliable \cite{Hu:2000ee}. 
We find that the dipole contribution leads to a small downward shift of the
spectrum at low values of $l$. The shift is large at low $l$ and roughly 1\% 
at $l$ of order 40.    Hence it is small but not negligible.  Some of the numerical values for B-mode power spectrum and corresponding fractional changes are given in Tab. (\ref{tabfrac1}). At higher values $l\geq100$, the fraction change becomes below $0.1\%$.
In Fig. (\ref{fig2}), on the left panel, we  note that $W_{11}^{l'}- W_{11}^{l'}$ is oscillatory and mainly lie in positive-value quadrant. Due to this asymmetry, dipole distribution could modify $C_{Bl}$. On other hand, $W_{11}^{l'}+ W_{11}^{l'}$ is oscillatory as well as symmetric. Therefore, we do not get any 
contribution to the $C_{El}$ due to the dipole distribution. Similarly, $W_{11}^{l'}$ forms a nearly symmetric pattern (see Fig. (\ref{fig3}))  and it does not give any change to $C_{Tl}$ in the standard weak lensing modification due to dipole anisotropy.

\section{Conclusions}\label{Sec5}
We generalized the formalism of weak lensing by adding dipole anisotropy
in the Newtonian potential. We estimated the effect of lensing on the CMB power spectrum in this model. We followed the approach of the flat sky approximation. Weak lensing mixes E and B mode and hence we necessarily find some 
contribution to $C_{Bl}$
due to weak lensing. The correction to $C_{Bl}$ is proportional to 
$\tilde C_{El}$. 
We observed that adding dipole anisotropy changes $C_{Bl}$ for the lower range of $l$. However, dipole anisotropy does not lead to an appreciable  
change in $C_{El}$, since the correction is much smaller than the leading order term $\tilde C_{El}$. We did not even observe any change in the standard weak lensing modification due to the dipole anisotropy, since $W_{11}^{l'}+ W_{11}^{l'}$ has a symmetric pattern. The same argument applies to $ C_{Tl}$.
 In the case of $C_{Bl}$ we obtain a correction since 
in this case the leading order $\tilde C_{Bl}$ is zero and $W_{11}^{l'}- W_{11}^{l'}$ are not symmetric. For this case we computed the maximum possible 
correction to 
$C_{Bl}$ by fixing the direction of dipole to be same as the
  direction of observation. A more reliable calculation would use the 
 spherical harmonics approach. Flat sky approximation and the spherical harmonics approach deviate below $l=10$ \cite{Hu:2000ee}. Thus, our calculation is reliable for $l>10$, where we still find a small correction to $C_{Bl}$.

\end{document}